\documentclass[reprint,aip,amsmath,amssymb,nofootinbib]{revtex4-1}
\usepackage{amssymb}
\usepackage{amssymb}
\usepackage{amssymb}
\usepackage{graphicx}
\usepackage{graphics}
\usepackage{amsmath}
\usepackage{placeins}
\usepackage{hyperref}
\usepackage[dvipsnames]{xcolor}
\usepackage{wasysym}
\usepackage{soul}
\usepackage{float}
\usepackage{array}
\usepackage{tabulary}
\usepackage{caption}
\usepackage{fancyhdr}

\begin{document}
	
	\title{FEgen (v.1): Field Emission Distribution Generator Freeware Based on Fowler-Nordheim Equation}
	
	\author{\firstname{Emily} \surname{Jevarjian}}
	\email{jevarji1@msu.edu}
	\affiliation{Department of Physics and Astronomy, Michigan State University, MI 48824, USA}
    \affiliation{Department of Electrical and Computer Engineering, Michigan State University, MI 48824, USA}
	\author{\firstname{Mitchell} \surname{Schneider}}
	\email{schne525@msu.edu}
	\affiliation{Department of Physics and Astronomy, Michigan State University, MI 48824, USA}
    \affiliation{Department of Electrical and Computer Engineering, Michigan State University, MI 48824, USA}
	\author{\firstname{Sergey V.} \surname{Baryshev}}
	\email{serbar@msu.edu}
	\affiliation{Department of Electrical and Computer Engineering, Michigan State University, MI 48824, USA}
	
	
	\begin{abstract}
	
	\end{abstract}
	
	
	\maketitle
	
	\section*{Introduction}
As field emitters are poised to become the preferred electron source for next-generation electron accelerators and other vacuum electronics microwave devices moving up in operating frequency for higher peak power rating and compactness,\cite{twt, JiaqiIEEE2018, Shao2019,THzgun,LewellenPRST2005} a computational toolbox must be developed to realistically model the particle dynamics. Unlike in photoemission, where an ultrashort high power density laser pulse is synchronized (in other words, phase matched) with an rf/microwave drive signal, electrons are generated by and interact with the rf/microwave drive cycle in a much wider phase window, regardless of whether a field emission cathode is operated in an ungated or gated fashion (by means of a physical gate electrode, harmonics mixing, or multicell gun design). An extended interaction phase window is of paramount importance to correctly reveal the longitudinal phase space of the resulting beam, which may promote delayed emission and secondary effects in the injector ultimately leading to beam loading, multipacting, and cathode field screening effects.
Compactness of a high frequency system (and the corresponding small emitting area of an electron source required to emit high charge) poses challenges in regard to correctly tracking and accounting for vacuum space charge effect and beam expansion/explosion. Currently, all in-demand beam tracking software that is capable of accounting for space charge effects, such as ASTRA,\cite{ASTRA} IMPACT-T,\cite{impact-t} or GPT,\cite{gpt} does not incorporate a particle distribution generator suitable for field emission analyses. There are costly PIC codes, VSim\cite{VSim} and Michelle\cite{michelle} for example, that account for the space charge effect yet contain field emission models likely based on the conventional Fowler-Nordheim equation. However, the exact mechanism of how Michelle and VSim determine their field emission distributions is proprietary.
This work is aimed to reengineer the generator function designed for ASTRA using Python, implement the Fowler-Nordheim equation to allow for observing temporal/phase (and thus complex longitudinal) beam processes, and make it a freeware as the ASTRA generator is only available as an executable file and has a number of application issues. We compare our results to those obtained using the ASTRA generator to verify our results for the coordinate-momentum space distributions. The generated distributions have the same format and structure as those obtained from ASTRA and can be directly translated into both ASTRA and GPT, the latter of which was used for our detailed application example. In particular, this work provides the ability to design and simulate transversely inherently shaped beams using array field emission cathodes providing new means for improving wakefield structure or plasma accelerators.\cite{beamlet} The paper is laid out as follows: Section I shows the momentum distribution, Section II shows the spatio-temporal distribution, and Section III shows a GPT application example for an L-band injector design. Section IV briefly discusses further developments for FEgen beyond the Fowler-Nordheim equation for future releases.

To download Python software package go to \url{https://github.com/schne525/FEgen}
	
	\section{\label{cha1}Momentum Distribution }
\begin{figure*}[htp]
		\includegraphics[width=13cm]{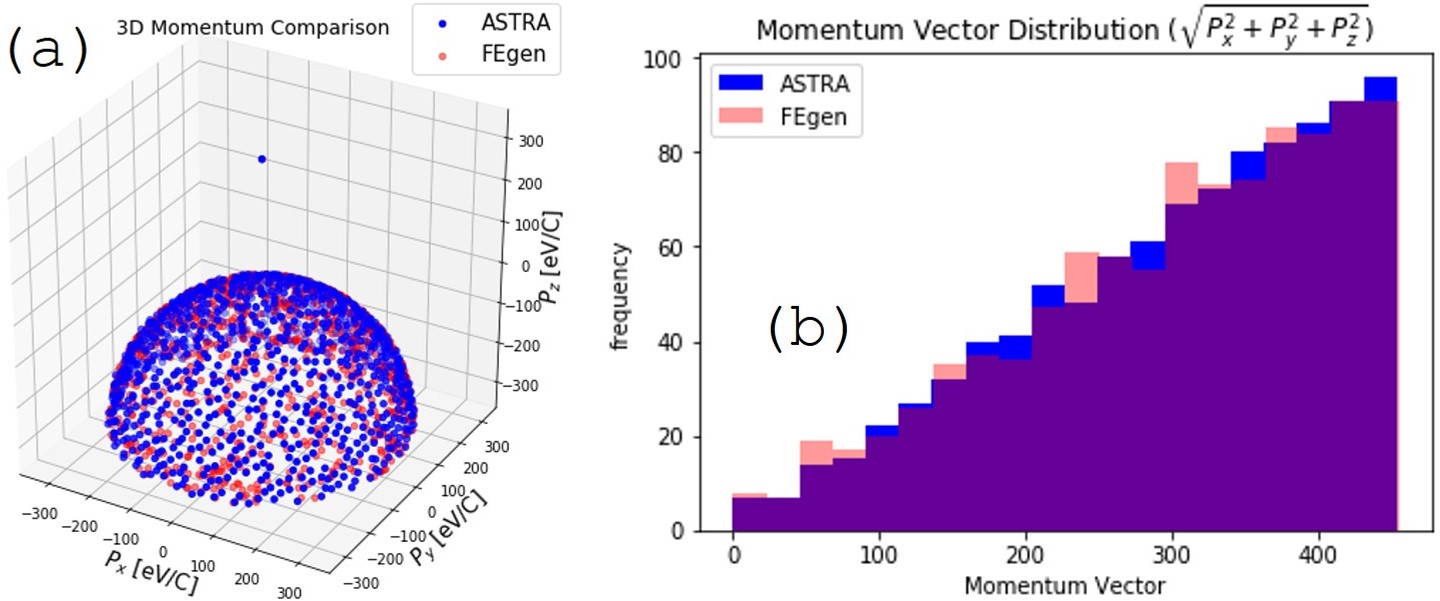}
		\caption{\label{f1} Apples-to-apples comparison between ASTRA and FEgen developed by the authors: (a) isotropic distribution in momentum space. The single point above the sphere is the test particle where all of the momentum is in the $z$ direction; (b) the distribution for the magnitude of the momentum vector.}
	\end{figure*}

The momentum distribution is based off an isotropic distribution which arises from the emittance on the cathode surface as the electrons tunnel through the barrier. This results in a momentum spread that is uniformly distributed over a half sphere, where the base is the surface of the cathode. This is referred to as an \textit{isotropic distribution}. To create the isotropic momentum distributions, first the maximum energy of each particle is calculated. Using Numpy’s random normal distribution function, a three-dimensional array of values with a normal distribution between -1 and 1 in each dimension is created to serve as unit vectors for each particle’s momentum in the $x$, $y$, and $z$ direction. To compare to ASTRA's generator, the $x$ and $y$ components of the unit vector array are multiplied by the maximum energy for a particle, creating a uniform kinetic energy distribution. Subsequently, this product is multiplied by a scaling factor which accounts for the uncertainty in the original $x$ and $y$ momentum distributions created by ASTRA. We found this scaling factor for uncertainty to be needed only in the $x$ and $y$ dimensions, and that the $z$ component of the momentum distribution mimicks that of ASTRA’s without an adjustment for uncertainty (see Fig.~\ref{f1}).

In Fig.~\ref{f1}, the absolute value of the $z$ dimension of the unit vector array is multiplied by the maximum energy for a particle, given that momentum in the $z$ direction must be positive. The magnitude of each particle’s momentum vector is then calculated and used to create an array of each particle’s energy which is then used in Scipy’s Statistical Kolmogorov-Smirnov test (kstest) to generate a $p$-value indicating the uniformity of the energy distribution by testing against Scipy’s Statistic Uniform distribution function. The given $p$-value is then compared to a significance level of 0.01. When the generated momentum arrays meet this significance condition, the momentum values are accepted and stored in the program to later be written into the output file along with the spatio-temporal components.
	
	\section{\label{cha2}Spatio-temporal Distribution }
\begin{figure*}[htp]
		\includegraphics[width=13cm]{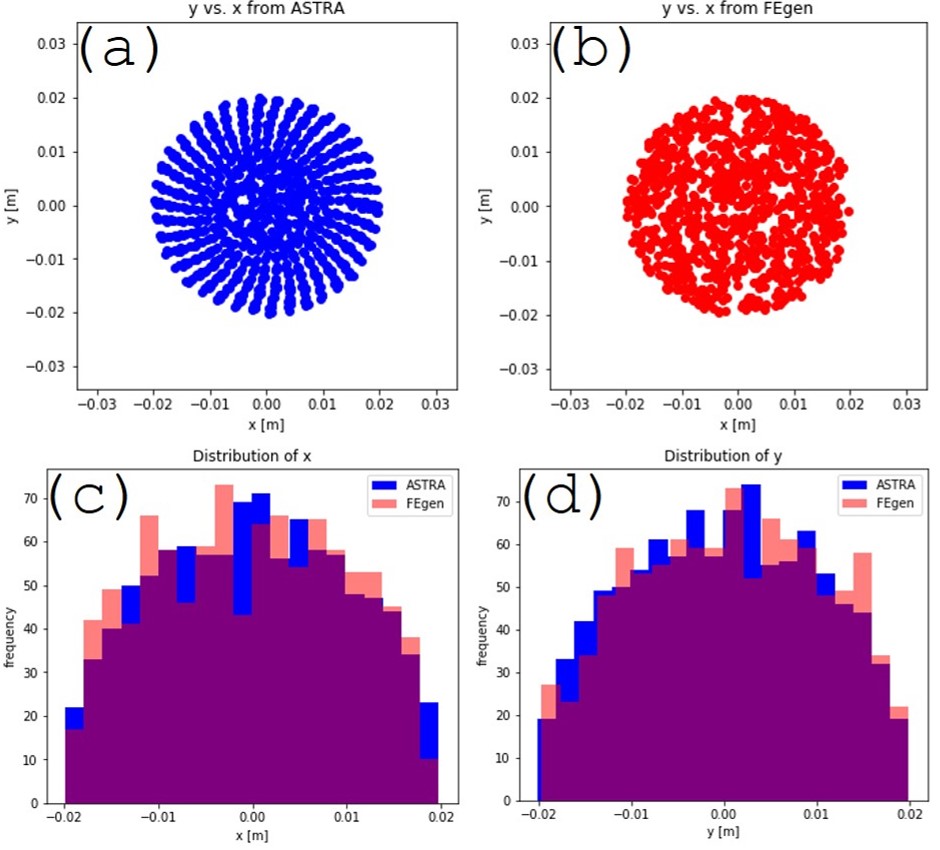}
		\caption{\label{f2} Radial distribution calculated using (a) ASTRA and (b) FEgen; (c) and (d) compare between $x$ and $y$ distributions obtained with ASTRA and FEgen.}
	\end{figure*}

	Fig.~\ref{f2}a illustrates how the radial distribution implanted in ASTRA resembles a spiral pattern, as if its random number generator uses the Fibonacci spiral with the ratio of $\pi^{-1}$. The spatial radial distribution is uniform over a given radius. For FEgen, Numpy’s random distribution was used to produce the radial distribution, as illustrated in Fig.~\ref{f2}a. This conceptual difference does not affect the resulting individual distributions of the $x$ and $y$ coordinates as clearly emphasized by Fig.~\ref{f2}c and ~\ref{f2}d.
	
\begin{figure*}[htp]
		\includegraphics[width=15cm]{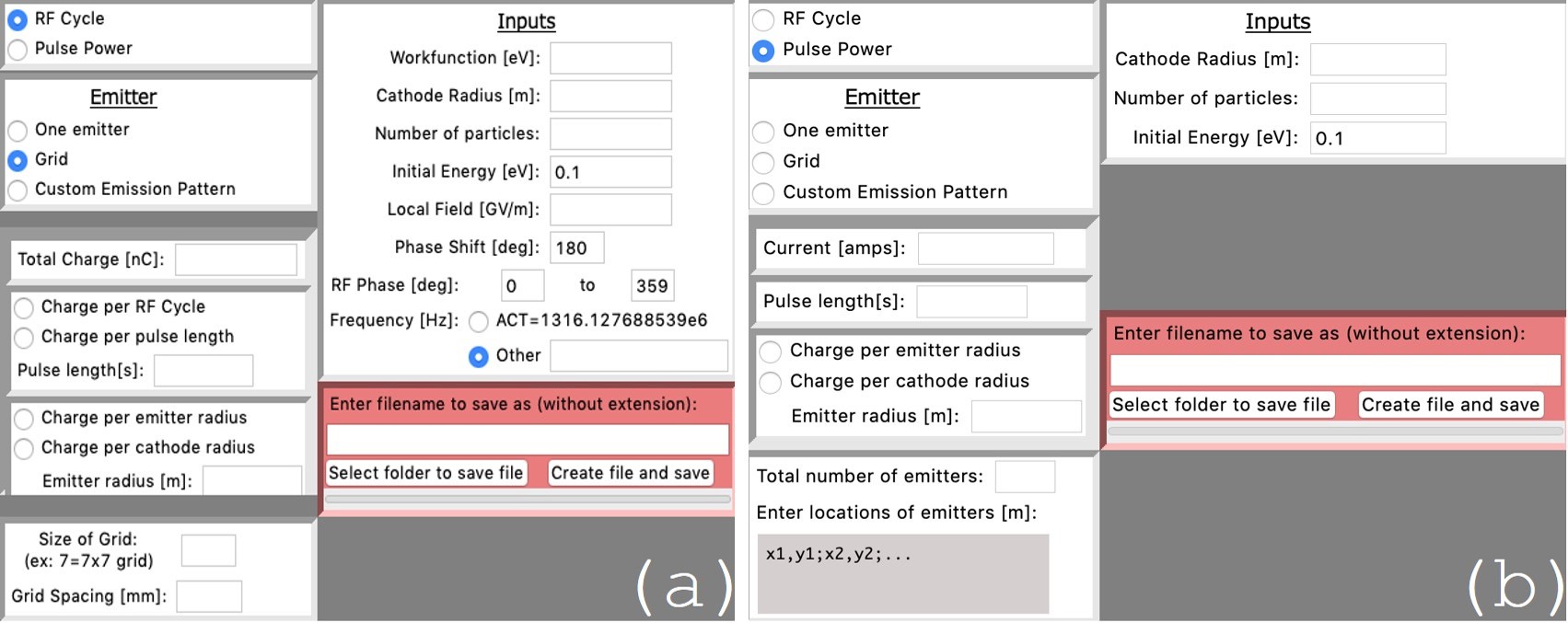}
		\caption{\label{f3} FEgen interface for initial particle distribution generation. Functionalities include 1) rf and dc pulsed power environments, ability to design 2) uniformly spaced grid of emitters and 3) custom grid of emitters. Note – Pulsed power function uses the pulse length of a dc system and the current to calculate the charge. As in any dc environment, the output current is constant with respect to time and does not follow the Gaussian-like distribution associated with the Fowler-Nordheim equation containing time varying electric field. FEgen interface functionality for (a) rf and (b) dc pulsed power environments.}
	\end{figure*}

	The FEgen has additional features such that, beyond having a single emitter, one can design a variety of emission patterns to simulate custom emitter arrays. A user can pick not only the radius of the emitter but also design an emitter grid and a custom pattern of emission points. An additional benefit is if the user knows the total charge of the beam or the total charge over the entire emission pattern region, FEgen can calculate then the charge for each emitter. This is useful in the case of simulating an emission grid of only a few emitters to represent a uniform emission which may have thousands of emitters on the cathode surface to maintain the ratio of emission area to charge to accurately simulate the space charge forces on the beam downstream. The interface of the initial particle distribution FEgen, containing all of the aforementioned features, is shown in Fig.~\ref{f3}.
	
	The temporal distribution is determined upon whether the field emission source is operated in a dc pulsed power (dc) or rf (ac) environment. In the dc environment, the field emission current is constant with time. Therefore, the temporal distribution follows a uniform distribution where the output charge is found by inputs for the pulse length and current.
	
\begin{figure}[htp]
		\includegraphics[width=8cm]{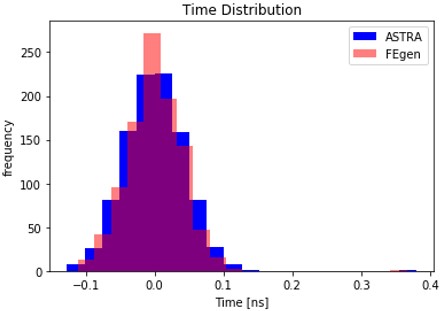}
		\caption{\label{f4} Comparison between ASTRA and FEgen for the temporal distribution.}
	\end{figure}

	In the rf environment, the Fowler-Nordheim equation is time-dependent. When averaged over an rf cycle, the Fowler-Nordheim equation is transformed into a form that reads\cite{JuwenSLAC1997}
	
	\begin{equation}\label{eq1}
	\begin{aligned}
	I_{F}(t)=&\frac{1.54\times 10^{-6}\times 10^{4.52\phi ^{-0.5}}A_{e}[\beta E_{c}(t)]^{2}}{\phi}\\
	&\times \exp[-\frac{6.53\times 10^{9}\phi ^{1.5}}{\beta E_{c}(t)}],\\
	\end{aligned}
	\end{equation}
\noindent here the external electric field is modeled as a time variant sinusoidal oscillation which is a result of only considering the longitudinal component. Eq.~\ref{eq1} is then fitted to a Gaussian distribution to determine the mean and standard deviation over the emission phase as specified by the input parameters (exemplified in Fig.~\ref{f4}).

Generally speaking, there is no intrinsic gating in field emission, and the current is allowed to emit over 360$^\circ$ of the rf cycle, and only electric field strength in Eq.~\ref{eq1} dictates when the emitting charge quenches. On the other hand, Eq.~\ref{eq1} is highly non-linear and it is therefore hypothesized that the emission phase window is much shorter than 360$^{\circ}$, often assumed\cite{LewellenPRST2005} to be equal to 60$^{\circ}$ ($\pm$30$^{\circ}$ around the rf cycle electric field crest). This is of course not a fundamentally defined threshold, and FEgen interface offers to input a specific rf phase where the emission to occur thus allowing, e.g., for finding best agreement between simulations and experimental data. The other important input parameters are the work function of the cathode material, initial energy spread at the cathode surface, the phase shift and frequency of the gun, and the local field on the cathode surface (the product of the applied field and the field enhancement factor $\beta$). The FEgen code was originally intended for the Argonne Cathode Teststand (ACT) where the default frequency is the L-band operational frequency of 1.316 GHz. The initial energy distribution at the cathode surface is defaulted to 0.1 eV, as for most materials the initial energy distribution is a fraction of an eV.\cite{TED} As this current model uses the Fowler-Nordheim equation, all particles are assumed to emit at $z$=0, which is the location of the cathode surface in the simulation given that the cathode surface is of a planar geometry. Even so, some useful results can be obtained when simulating a field emission cathode which consists of a specifically designed patterned array of emitting tips – this is illustrated in the next section.

\section{\label{cha3}Application Example}

Currently, photoemission sources are used to produce transversely shaped electron beamlets by means of transversely shaped lasers or masks on the cathode surface to construct a desired pattern. A transversely shaped laser pulse using additional optics complicates overall system design and may diminish the overall power of the laser. However, a field emission source specifically engineered with a specific pattern/array of emitters enables new means for direct transverse multi-beamlet generation in a high efficiency fashion. Emittance exchange techniques further allow for conversing transversely shaped beams into longitudinal bunch trains critical for the development of the next generation wakefield structure or plasma accelerators. One example of recent successful demonstration of transversely shaped beams is described in Ref.\cite{beamlet}: specifically, it was demonstrated that using a grid of pyramidal nanodiamond emitters a transversely shaped beam could be transported nearly intact $\sim$2 m downstream from the injector. The presented application example shows FEgen’s capability to help reproduce this experimental result with high fidelity by tracking particles in GPT.

\begin{figure}[htp]
		\includegraphics[width=8cm]{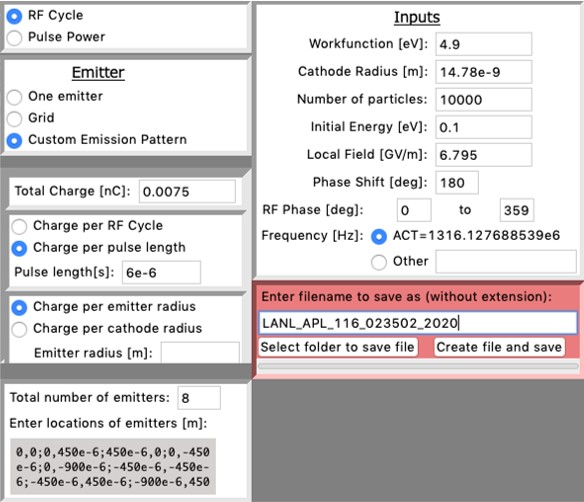}
		\caption{\label{f5} Parameters set for creating initial field emission distribution and the custom emission pattern replicating one described in Ref.\cite{beamlet}.}
	\end{figure}

Fig.~\ref{f5} is a screenshot of FEgen’s interface for this simulation based upon the parameters presented in Ref.\cite{beamlet}. The pyramids were simulated as radially uniform planar emitters with a radius equal to 15 nm spaced 450 $\mu$m away from each other. The total charge is the charge per each of 8 emitters (where the total beam charge reported in Ref.\cite{beamlet} was 60 pC), acquired over an rf pulse length of 6 $\mu$s. When the option for charge per pulse length is selected in FEgen, it determines the charge in a single rf cycle by determining the number of rf cycles within the rf pulse length based upon the gun operating frequency entered (1.316 GHz in this case). The phase shift is given as a default value of 180$^\circ$ to ensure that the reference particle is at the field crest at the time of emission. This value comes from the phase shift of the field by 90$^\circ$ due to cosine field used in Superfish and sine field used by ASTRA and GPT. The field crest is found at an additional shift of 90$^\circ$, hence the total phase shift is set at a default of 180$^\circ$. The local field represents the product of the applied field times the field enhancement factor $\beta$, which was found to be 450, yielding a local field of 6.795 GV/m (given an operating applied field of 15.1 MV/m.\cite{beamlet}) The cathode radius was determined using the effective emission area calculated from the Fowler-Nordheim equation with a total effective emission area of 5,490 nm$^2$ for all eight emitters. This led to the radius of the emitter for each of the pyramids being 14.78 nm which closely matched with the actual physical size of nano-diamond pyramid tips. (See Appendix A for more details on input parameters.) 

\begin{figure}[htp]
		\includegraphics[width=8cm]{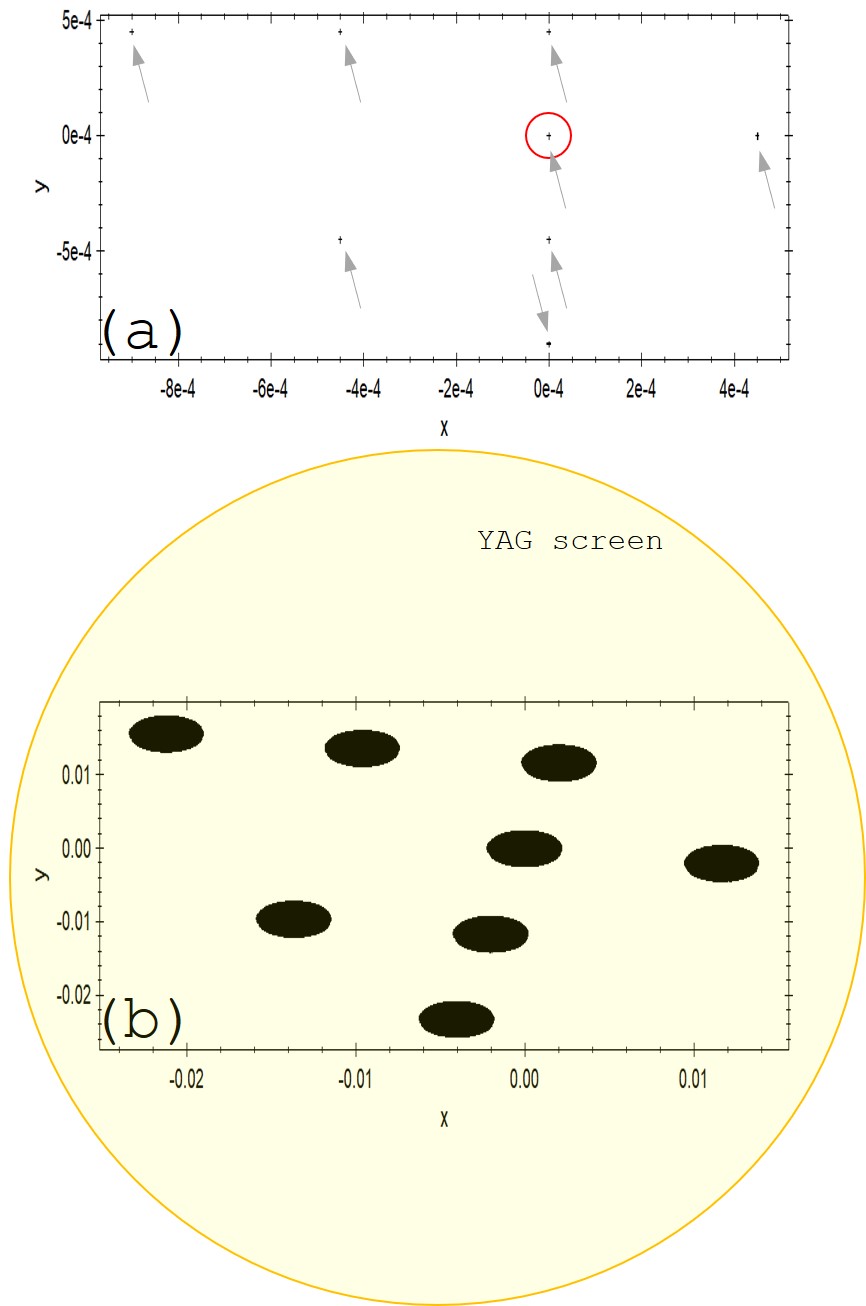}
		\caption{\label{f6} (a) Initial emission pattern on the cathode surface. Indicated by an arrow is the location of each emitter with a initial radius of 14.78 nm. The circle represents the emitter that was denoted as being placed at the origin. This emitter was selected to be at the origin because it was the brightest emitter in Ref.\cite{beamlet}, indicating that that emitter is generated on-axis; (b) Transverse electron beamlet distribution at 2.54 m downstream from the cathode ($z$=0). YAG screen has radius of 0.0254 m, therefore the entire emission pattern would be projected onto the YAG screen without loss.}
	\end{figure}

This initial distribution was then imported into GPT together with the Superfish field maps for both the gun and the imaging and focusing solenoids implemented on the ACT beamline. The simulation was conducted using a gun gradient of 15.1 MV/m and the solenoid settings were optimized to transport the pattern beam downstream. Fig.~\ref{f6}a illustrates the designer emission pattern on the cathode surface as plotted in GPT at $z$=0 and $t$=0, while Fig.~\ref{f6}b shows the transverse electron distribution at a location along the beamline that would correspond to the position of the imaging screen YAG3 ($z$=2.54 m). This results demonstrates and proves computationally that the transversely shaped beam, once generated, can be transported downstream for a long distance. Another conclusion from this simulation is that it was primarily possible to achieve thanks to relatively low space charge effect. When coupled with beam tracking software, this result further proves that a useful means was created to easily design custom emission patterns allowing for rapid R\&D of intrinsically shaped beams, finding beamline settings to transport them, and comparing to experiment.

	\section{\label{cha4}Outlook for Future Releases}
	A limitation from the application example is that the designed field emitter is a pyramid; however, at this current time only the very tip of the emitter ($z$$\approx$0) was considered. It is known from experimental measurements that the emitter first emits from the tip and then, as the external field increases, more of the side walls begin to emit, as the turn-on field for the pyramid is a function of $z$ and the geometry of the tip emitter. Thus, the next step is to expand this work with Fowler-Nordheim equations for non-planar geometries to model such a physics accurately. The potential distribution implementation over curved surfaces will strongly affect the momentum and the spatio-temporal distributions.
	
	The next generation of field emission sources, as shown in the application example, are most likely to be semiconductor materials, such as those made of carbon nanotubes and diamond materials.\cite{afrl,SergeyAPL2014, Taha, nanoD, piot1, piot2, FNdeviation, ChubenkoJAP2019} Recent results have shown a divergence from the classical Fowler-Nordheim conditions,\cite{FNdeviation, Taha, ChubenkoACS2017, ChubenkoJAP2019} which is attributed to the current model failing to account for an emitter being a semiconductor. Future work will implement the semiconductor effects using the Stratton-Baskin-Lvov-Fursey formalism,\cite{ChubenkoJAP2019} expanding into the rf environment, temperature and patchy/varying work function effects. Such implementations will be completed for future releases. The goal in mind is to modify FEgen to account for these effects and to ultimately understand the new emission physics that can be seen through experimental studies.
	
		\begin{acknowledgments}
		We would like to thank Dr. Zenghai Li of SLAC and Jiahang Shao of ANL for illuminating discussions.
		This work was supported by the US Department of Energy, Office of Science, High Energy Physics under Cooperative Agreement award No. DE-SC0018362 and Michigan State University, the College of Engineering, Michigan State University, under the Global Impact Initiative. This material is also based upon work supported by the U.S. Department of Energy, Office of Science, Office of High Energy Physics under Award No. DE-SC0020429.
	\end{acknowledgments}
	\pagebreak
	\section*{References}
	\bibliography{UNCD_ref}
	
\begin{table*}[!h]
\section*{Appendix A}
\caption*{For all entry of parameters using scientific notation, use $e$ notation to prevent errors when inputs are processed (e.g. enter 1.2e3 for 1200, enter 1.2e-6 for 0.0000012). Details on entering parameters are provided in the table below.}
\begin{center}
{\renewcommand{\arraystretch}{2}
\begin{tabular}{ m{3cm} | m{12cm}| m{1cm} } 
\hline \hline
 \textbf{ Parameter} & \textbf{Details} & \textbf{Unit} \\
  \hline
  One Emitter & Select for a single emitter & -- \\
  
  \hline
  
  Emitter Grid & Select to create a square grid of emitters. See parameters for Size of Grid and Grid Spacing for details on specifying grid size & -- \\
  
  \hline
  
  Custom Emission Pattern & Select to create a custom pattern of emitters. See parameter for Locations of Emitters for details on specifying emitter locations & -- \\
  
  \hline
  
 Size of Grid & The length of one side of a square grid of emitters (e.g. Size of Grid = 7 will create a 7$\times$7 grid of 49 total emitters) & -- \\
 
  \hline
  
 Grid Spacing & The distance between emitters when grid of emitters is selected & mm \\
 
   \hline
  
 Locations of emitters & The location of each emitter in meters. See note at beginning of Appendix A on how to enter values using scientific notation. Ensure that the number of emitters entered coincides with the number of ($x$, $y$) pairs given. To enter locations, separate respective $x$ and $y$ with a comma and place a semicolon between ($x$, $y$) pairs. Do not place a semicolon after the last ($x$, $y$) pair. Straying from this format will result in an error. See Fig.~\ref{f5} as an example & m \\
  
    \hline
  
  Current & For the pulsed power option, current and pulse length are used to calculate charge $q=I\times\tau_{dc}$ where $\tau_{dc}$ is the dc pulse length & A \\
  
  \hline
  
  Total Charge & Charge per emitter (e.g. for 1 nC charge over 10 emitters, one would enter 0.1 nC) & nC \\
  
   \hline
  
  Charge per rf Cycle & Select for charge in single rf cycle & nC \\
  
    \hline
  
  Charge per Pulse Length & Select for charge over multiple rf cycles. Pulse length and frequency are used to calculate charge per cycle $q=\frac{q_{total}}{f\times\tau_{rf}}$ where $f$ is the operating frequency and $\tau_{rf}$ is the rf pulse length & nC \\
  
   \hline
  
  Charge per Emitter Radius & Select to scale the total charge to the emitter radius & nC \\
  
    \hline
  
  Charge per Cathode Radius & Select to scale the total charge to a smaller cathode radius $q=q_{total}(\frac{r_{emitter}}{r_{cathode}})^2$ & nC \\
    
    \hline
  
  Initial Energy & Initial energy distribution at cathode surface (default is 0.1 eV) & eV \\
    
    \hline
  
  Local Field & Equal to $\beta$ times applied field where $\beta$ is the field enhancement factor & GV/m \\
    
    \hline
  
  Phase shift & Accounts for phase shift of the ac field. Default value is 180$^\circ$ due to a 90$^\circ$ phase shift between the cosine and sine field (used in Superfish and ASTRA/GPT, respectively) and an additional 90$^\circ$ shift such that the reference particle is at the peak field at the time of emission & degree \\
      
    \hline
  
 rf Phase & Phase window of the rf cycle. Default is 360$^\circ$, phase window can be specified upon entry & degree \\
      
    \hline
  
  Frequency & Default option provided is for the Argonne Cathode Teststand (ACT) for which FEgen was originally intended where the default frequency is the L-band operational frequency of the ACT. Alternate frequencies can be entered by user & Hz \\
      
\hline \hline
\end{tabular}}
\end{center}
\end{table*}
	
\end{document}